# Tuning PID and PI$^\lambda$D$^\delta$ Controllers using the Integral Time Absolute Error Criterion

Deepyaman Maiti, Ayan Acharya, Mithun Chakraborty, Amit Konar
Dept. of Electronics and Telecommunication Engineering
Jadavpur University
Kolkata, India
deepyamanmaiti@gmail.com, masterayan@gmail.com, mithun.chakra08@gmail.com, konaramit@yahoo.co.in

Ramadoss Janarthanan
Dept. of Information Technology
Jaya Engineering College
Chennai, India
srmjana_73@yahoo.com

*Abstract*—Particle swarm optimization (PSO) is extensively used for real parameter optimization in diverse fields of study. This paper describes an application of PSO to the problem of designing a fractional-order proportional-integral-derivative (PI$^\lambda$D$^\delta$) controller whose parameters comprise proportionality constant, integral constant, derivative constant, integral order ($\lambda$) and derivative order ($\delta$). The presence of five optimizable parameters makes the task of designing a PI$^\lambda$D$^\delta$ controller more challenging than conventional PID controller design. Our design method focuses on minimizing the Integral Time Absolute Error (ITAE) criterion. The digital realization of the deigned system utilizes the Tustin operator-based continued fraction expansion scheme. We carry out a simulation that illustrates the effectiveness of the proposed approach especially for realizing fractional-order plants. This paper also attempts to study the behavior of fractional PID controller vis-à-vis that of its integer-order counterpart and demonstrates the superiority of the former to the latter.

*Keywords*—Continued fraction expansion, fractional calculus, ITAE criterion, particle swarm optimization, PID and PI$^\lambda$D$^\delta$ controllers

## I. Introduction

Dynamic systems based on fractional order calculus [1] have been a subject of extensive research in recent years since the proposition of the concept of the fractional-order PI$^\lambda$D$^\delta$ controllers and the demonstration of their effectiveness in actuating desired fractional order system responses by Podlubny [2].

In the current literature on control engineering, one can find quite a few recent works in this direction as well as schemes for digital and hardware realizations of such systems in [3] – [6], a frequency domain approach based on expected crossover frequency and phase margin for the same controller design by Vinagre et al [7], a method based on the pole distribution of the characteristic equation in the complex plane by Petras [8]. Dorcak et al [9] propounded a state space design approach based on feedback pole placement. It is also possible to synthesize the fractional controller cascading a proper fractional unit to an integer order controller [10].

For many decades, proportional - integral - derivative (PID) controllers have been very popular in industries for process control applications. Their merit consists in simplicity of design and good performance, such as low percentage overshoot and small settling time (which is essential for slow industrial processes). Owing to the paramount importance of PID controllers, continuous efforts are being made to improve their quality and robustness.

An elegant way of enhancing the performance of PID controllers is to use *fractional-order controllers* where the I- and D-actions have, in general, non-integer orders. In a PI$^\lambda$D$^\delta$ controller, besides the proportional, integral and derivative constants, denoted by $K_p$, $T_i$ and $T_d$ respectively, we have two more adjustable parameters: the powers of $s$ in integral and derivative actions, viz. $-\lambda$ and $\delta$ respectively. As such, this type of controller has a wider scope of design, while retaining the advantages of classical PID controllers. Finding the appropriate settings of the values of the five parameters $\{K_p, T_i, T_d, \lambda, \delta\}$ to achieve optimal performance for a given plant, as per user specifications, thus calls for real parameter optimization on the five-dimensional space. Our design method focuses on minimizing the ITAE criterion.

Classical optimization techniques are not applicable here because of the roughness of the multidimensional objective function surface. We, therefore, use a derivative-free optimization technique — particle swarm optimization (PSO) originally devised by Kennedy and Eberhart [11]. It draws inspiration from the intelligent, collective behavior of a *swarm* of social insects (particularly bees) foraging for food together. PSO and (subsequent modifications thereof) are highly regarded in research communities due to its combination of simplicity (in terms of its implementation), low computational cost and remarkable efficacy [12].

The remainder of the paper is organized as follows. Section II presents the fundamentals of fractional calculus, PID controllers of both integral and fractional orders and, finally, state-of-the-art methods for discretizing control systems. Section III seeks to review PSO algorithm while section IV details our controller synthesis procedure. Section V concludes the paper.

## II. INTEGER ORDER PID AND FRACTIONAL ORDER $PI^\lambda D^\delta$ CONTROLLERS

The dynamics of fractional order control systems are described by fractional order differential equations. Clearly, in order to grasp the significance of such systems, an understanding of the *theory of fractional calculus* is necessary.

### A. Theory of Fractional Calculus

At first, we generalize the differential and integral operators into one fundamental operator $_aD_t^\alpha$ where:

$$_aD_t^\alpha = \frac{d^\alpha}{dt^\alpha} \text{ for } \Re(\alpha) > 0;$$

$$= 1 \text{ for } \Re(\alpha) = 0;$$

$$= \int_a^t (d\tau)^{-\alpha} \text{ for } \Re(\alpha) > 0.$$

$\Re(\alpha)$ denotes the real part of $\alpha$ which is, in general, a complex quantity. For our purpose, $\alpha$ is purely real.

The two definitions used for fractional differintegral are the Riemann-Liouville definition and the Grunwald-Letnikov definition [1]. The Grunwald-Letnikov definition is:

$$_aD_t^\alpha f(t) = \lim_{h \to 0} \frac{1}{h^\alpha} \sum_{j=0}^{\left[\frac{t-a}{h}\right]} (-1)^j \binom{\alpha}{j} f(t - jh) \quad (1)$$

Derived from the Grunwald-Letnikov definition, the numerical calculation formula of fractional derivative can be achieved as:

$$_{t-L}D_t^\alpha f(t) \approx h^{-\alpha} \sum_{j=0}^{[L/T]} b_j f(t - jh) \quad (2)$$

L is the length of memory. T, the sampling time always replaces the time increment h during approximation. The weighting coefficients $b_j$ can be calculated recursively by:

$$b_0 = 1, b_j = \left(1 - \frac{1+\alpha}{j}\right) b_{j-1}, (j \geq 1). \quad (3)$$

### B. Basic Concept of PID and $PI^\lambda D^\delta$ Controllers

A PID controller is essentially a generic closed-loop feedback mechanism.

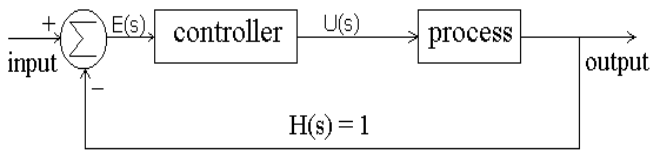

Figure 1. Unity feedback closed loop system

Its working principle is that it monitors the error between a measured *process variable* and a desired *set point*; from this error, a corrective signal is computed and is eventually fed back to the input side to adjust the process accordingly.

The differential equation of the PID controller is:

$$u(t) = K_p e(t) + T_i D^{-1} e(t) + T_d D e(t) \quad (4)$$

Thus, the PID controller algorithm is described by a *weighted sum* of three time functions where the three distinct weights are: the proportional gain ($K_p$) that determines the influence of the present error-value on the control mechanism, the integral gain ($T_i$) that decides the reaction based on the area under the error-time curve upto the present point and the derivative gain ($T_d$) that accounts for the extent of the reaction to the rate of change of the error with time. Thus, the superposition of these three actions constitutes the mechanism for adjustment of plant performance.

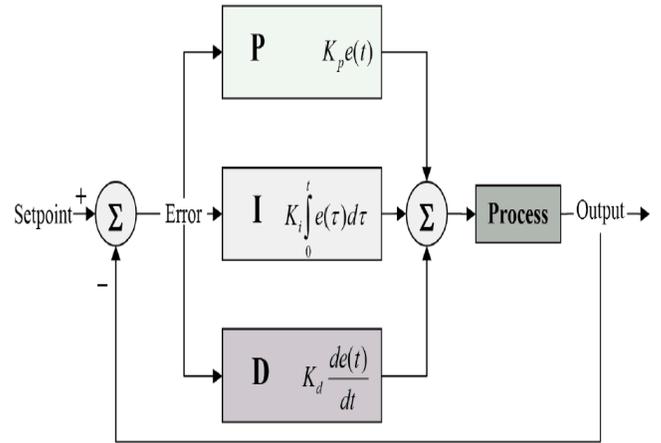

Figure 2. Generic closed loop control system with a PID controller

The differential equation of the $PI^\lambda D^\delta$ controller is described by [2]:

$$u(t) = K_p e(t) + T_i D^{-\lambda} e(t) + T_d D^\delta e(t) \quad (5)$$

The continuous transfer function of the $PI^\lambda D^\delta$ controller is obtained through Laplace transform as:

$$G_c(s) = K_p + T_i s^{-\lambda} + T_d s^\delta \quad (6)$$

After the introduction of this definition, it is easily seen that classical types of PID controllers such as integral order PID, PI or PD become special cases of the most general fractional order PID controller. In other words, the $PI^\lambda D^\delta$ controller *expands* the integer-order PID controller *from point to plane*, as shown in Fig. 2, thereby adding flexibility to controller design and allowing us to control our real world processes more accurately but only at the cost of increased design complexity. This is, however, not at all a heavy price paid for the benefits obtained.

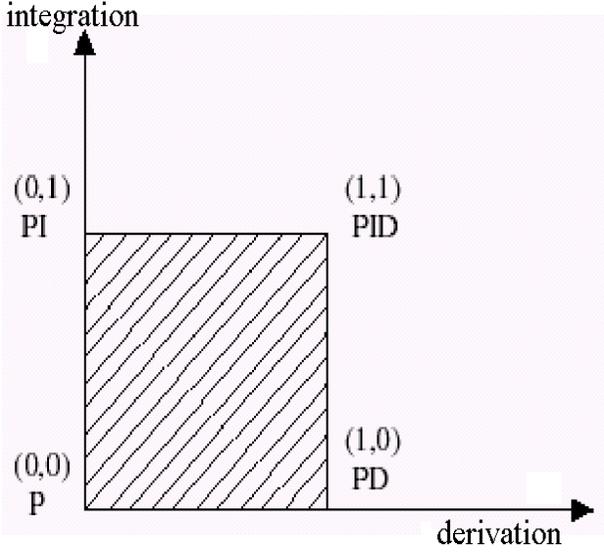

Figure 3. Generalization of the PID controller from point to plane

## C. Overview of Discretization Methods for Control Systems

For a perfect realization of fractional order controllers, all the past inputs should be retained in the memory. One can make use of the generating function $s = \omega(z^{-1})$ to transform the fractional order differentiator $s^r$ from s domain to z space.

Well-known $s \rightarrow z$ schemes are Euler and Tustin method. The coefficients of the approximation equations for fractional calculus may be obtained by considering the Tustin operator as generating function:

$$s^{\pm r} = (\omega(z^{-1}))^{\pm r} = \left(\frac{2}{T}\frac{1-z^{-1}}{1+z^{-1}}\right)^{\pm r} \quad (7)$$

and then perform the continued fraction expansion. The discretized result is:

$$Z\{D^{\pm r}x(t)\} = CFE\left\{\left(\frac{2}{T}\frac{1-z^{-1}}{1+z^{-1}}\right)^{\pm r}\right\}X(z)$$

$$\approx \left(\frac{2}{T}\right)^{\pm r}\frac{P_p(z^{-1})}{Q_q(z^{-1})}X(z) \quad (8)$$

where $CFE\{u\}$ denotes the continued fraction expansion of u; p and q are the orders of the approximation; $P_p$ and $Q_q$ are polynomials of degrees p and q respectively in the variable $z^{-1}$. Normally, we can set p = q = n. T is the sampling period. The general expression for numerator $P_p(z^{-1})$ and denominator $Q_q(z^{-1})$ of $D^{\pm r}(z)$ is summarized in table 1 for p = q = 1, 3, 5.

TABLE I. EXPRESSIONS FOR THE NUMERATOR AND DENOMINATOR POLYNOMIALS IN THE CFE

| p = q | $P_p(z^{-1})$ for k = 1 and $Q_q(z^{-1})$ for k = 0 |
|---|---|
| 1 | $(-1)^k z^{-1}r + 1$ |
| 3 | $(-1)^k(r^3 - 4r)z^{-3} +$ $(6r^2 - 9)z^{-2} +$ $(-1)^k 15z^{-1}r + 15$ |
| 5 | $(-1)^k(r^5 - 20r^3 + 64r)z^{-5} +$ $(-195r^2 + 15r^4 + 225)z^{-4} +$ $(-1)^k(105r^3 - 735r)z^{-3} +$ $(420r^2 - 1050)z^{-2} +$ $(-1)^k 945z^{-1}r + 945$ |

The $PI^\lambda D^\delta$ controller can be approximated using discretization methods, which is given by:

$$G_c(z) = K_p + T_i w_i(z) + T_d w_d(z) \quad (9)$$

where $w_i(z)$ is the discrete approximation equation of fractional order integral $s^{-\lambda}$, $w_d(z)$ is the discrete approximation equation of $s^\delta$.

The integer order controller can likewise be approximated in the z-domain by setting $\lambda = \delta = 1$.

## III. REVIEW ON PARTICLE SWARM OPTIMIZATION

The optimization problem consists in determining the *global optimum* (in our case, minimum) of a continuous real-valued function of *n* independent variables $x_1, x_2, x_3, ..., x_n$, mathematically represented as $f(\vec{X})$, where $\vec{X} = (x_1, x_2, x_3, ..., x_n)$ is called the *parameter vector*. Then the task of any optimization algorithm reduces to searching the n-dimensional hyperspace to locate a particular point with position-vector $\vec{X}_0$ such that $f(\vec{X}_0)$ is the global optimum of $f(\vec{X})$.

PSO [11] - [14] is in principle a multi-agent parallel search technique. We begin with a *population* or *swarm* consisting of a convenient number, say *m*, of *particles* — conceptual entities that "fly" through the multi-dimensional search space as the algorithm progresses through discrete (unit) time-steps $t$ = 0, 1, 2, ..., the population-size *m* remaining constant.

In the standard PSO algorithm, each particle *P* has two state variables: its current position $\vec{X}_i(t) = [X_{i,1}(t), X_{i,2}(t), ..., X_{i,n}(t)]$ and its current velocity $\vec{V}_i(t) = [V_{i,1}(t), V_{i,2}(t), ..., V_{i,n}(t)]$,

$i=1,2,\ldots,m$. The position vector of each particle with respect to the origin of the search space represents a *candidate solution* of the search problem. Each particle also has a small memory comprising its *personal best* position experienced so far, denoted by $\vec{p}_i(t)$ and the *global best* position found so far, denoted by $\vec{g}(t)$. Here, one position is considered *better* than another if the former gives a *lower* value of the objective function, also called the *fitness function* in this context, than the latter.

For each particle, each component $X_{i,j}(0)$ of the initial position vector is selected at random from a predetermined search range $[X_j^L, X_j^U]$, while each velocity component is initialized by choosing at random from the interval $[-V_{j\max}, V_{j\max}]$, where $V_{j\max}$ is the maximum possible magnitude of velocity of any particle in the $j$th dimension, $j = 1, 2, \ldots, n$, $i = 1, 2, \ldots, m$; the initial settings for $\vec{p}_i(t)$ and $\vec{g}(t)$ are taken as $\vec{p}_i(0) = \vec{X}_i(0), \vec{g}(0) = \vec{X}_k(0)$ such that $f(\vec{X}_k(0)) \leq f(\vec{X}_i(0)) \forall i$.

After the particles are initialized, the iterative optimization process begins, where the positions and velocities of all the particles are updated by the following recursive equations (10), (11). The equations are presented for the $j$th dimension of the position and velocity of the $i$th particle.

$$v_{id}(t+1) = \omega v_{id}(t) + c_1 \cdot \varphi_1 \cdot (p_{id}(t) - x_{id}(t)) + c_2 \cdot \varphi_2 \cdot (p_{gd}(t) - x_{id}(t)) \quad (10)$$

$$x_{id}(t+1) = x_{id}(t) + v_{id}(t+1) \quad (11)$$

where the algorithmic parameters are defined as :

- $\omega$ : the time-decreasing inertial weight factor designed by Shi and Eberhart [13],
- $C_1, C_2$ : two constant multipliers called *self confidence* and *swarm confidence* respectively,
- $\varphi_1, \varphi_2$ : two uniformly distributed random numbers.

The iterations are allowed to go on for a certain pre-determined number of time-steps (*maxiter*), or until the fitness of the best particle at a certain time-step is better than a pre-defined value (*tolerance*).

On termination of the algorithm, most of the parameter vectors are expected to converge to a small region around the required global optimum of the search space. The fittest vector of the final population is taken as a possible solution to the problem.

## IV. DESIGN OF THE CONTROLLER

Optimization by PSO consists of designing the optimization goal, i.e. the fitness function and then encoding the parameters to be searched. The PSO algorithm runs until the stop condition is satisfied. The best particle's position gives the optimized parameters.

### A. The Parameters to be Optimized

The $PI^\lambda D^\delta$ controller has five unknown parameters to be tuned, viz. $\{K_p, T_i, T_d, \lambda, \delta\}$. Hence the present problem of controller tuning can be solved by an application of the PSO algorithm for optimization on a five-dimensional solution space, each particle having a five-dimensional position and velocity vector.

For tuning the integer order PID controller, the solution space will be three-dimensional, the three dimensions being the three parameters of the controller, viz. $\{K_p, T_i, T_d\}$.

### B. PSO Factors

Number of PSO particles in the population is 10.

The inertia factor $\omega$ decreases linearly from 0.9 to 0.4, $c_1 = c_2 = 1.4$.

We used the PSO dynamics [13] for the experiments in this article.

The initialization ranges and the limits on positions and velocities of the parameters are summarized.

TABLE II. INTEGRAL ORDER CONTROLLER: RANGES OF PARTICLES

| Parameters | Position Vector | | Velocity Vector | |
|---|---|---|---|---|
| | Initialize | Limit | Initialize | Limit |
| $K_p, T_i, T_d$ | 0 to 500 | 0 to 500 | -1 to +1 | no limits |

TABLE III. FRACTIONAL ORDER CONTROLLER: RANGES OF PARTICLES

| Parameters | Position Vector | | Velocity Vector | |
|---|---|---|---|---|
| | Initialize | Limit | Initialize | Limit |
| $K_p, T_i, T_d$ | 0 to 500 | 0 to 500 | -1 to +1 | no limits |
| $\lambda, \delta$ | 0 to 2 | 0 to 2 | -1 to +1 | |

### C. Fitness Function

As already mentioned, the fitness function to be minimized is the ITAE performance criterion. The integral of the absolute magnitude of error (ITAE) criterion is defined as $ITAE = \int_0^T t|e(t)|dt$. The ITAE performance index has the advantages of producing smaller overshoots and oscillations than the IAE (integral of the absolute error) or the ISE (integral square error) performance indices. In addition, it is the most sensitive of the three, i.e. it has the best selectivity. The ITSE (integral time-square error) index is somewhat less sensitive and is not comfortable computationally [15], [16]. Since it is not practicable to integrate up to infinity, the convention is to choose a value of T sufficiently large so that e(t) for t > T is negligible. We used T = 10 seconds.

### D. Stop Criterion

The stop criterion used was the one that defines the maximum number of generations to be produced. We used 100 generations.

*E. Results*

The control objective has the transfer function $\dfrac{1}{0.8s^{2.2} + 0.5s^{0.9} + 1}$. A sampling period of 0.01 seconds was used. The reference input is the unit step: $R(s) = \dfrac{1}{s}$.

After the stop criterion is met, i.e. after 100 runs of the PSO algorithm, the position vector of the best particle gives the optimized parameters of the fractional order controller as: $K_p = 325.9739$, $T_i = 303.3286$, $T_d = 389.4627$, $\lambda = 0.6022$, $\delta = 1.6188$. The fitness of the best particle is $4.9094 \times 10^{-4}$. This, evidently, is the value of the ITAE index.

The optimized parameters of the integral order controller are $K_p = 47.9222$, $T_i = 29.6641$, $T_d = 449.1112$. The fitness of the best particle is $0.0979$. The time responses as well as the variations of best fitness indicate the superiority of the fractional order controller over the integer order one.

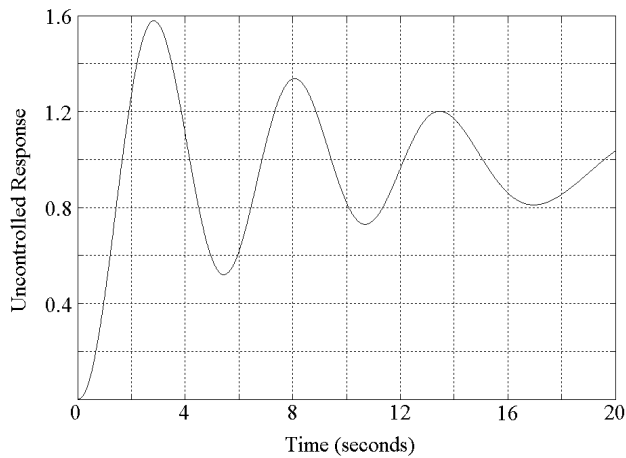

Figure 4.  Open loop step response of plant only

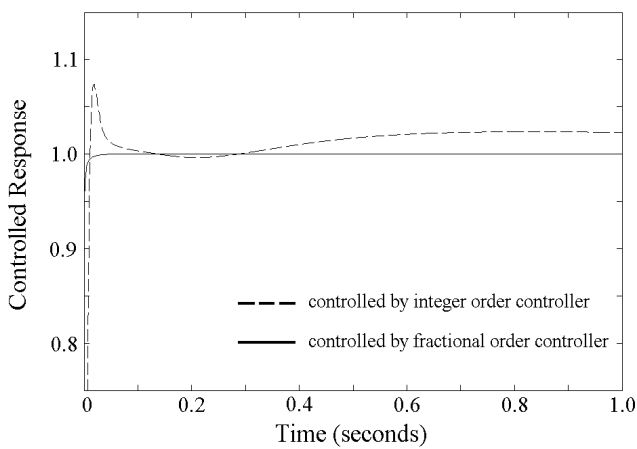

Figure 5.  Controlled closed loop response of the system

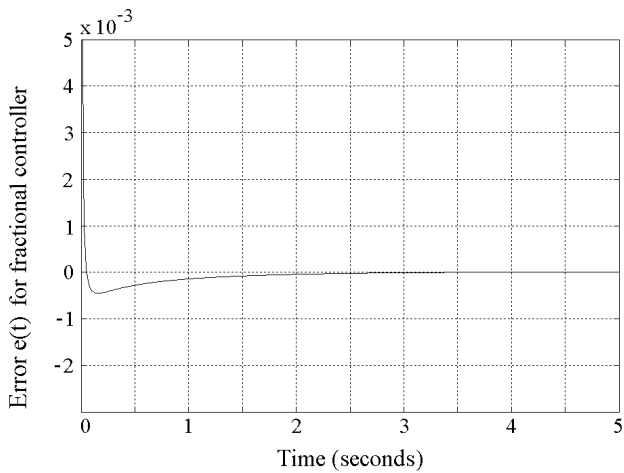

Figure 6.  Error waveform e(t) = r(t) – c(t) for plant controlled by fractional order controller

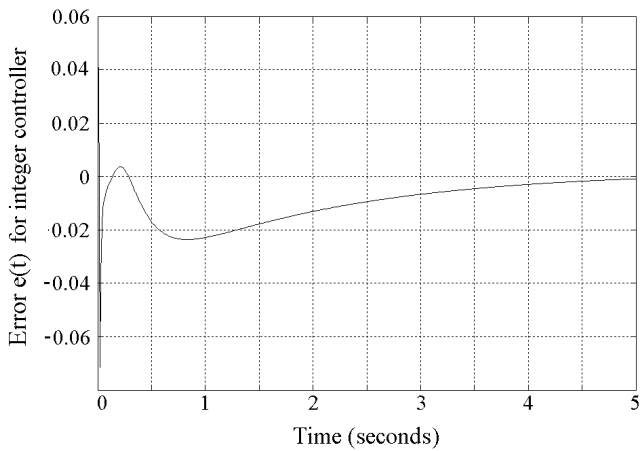

Figure 7.  Error waveform e(t) = r(t) – c(t) for plant controlled by integer order controller

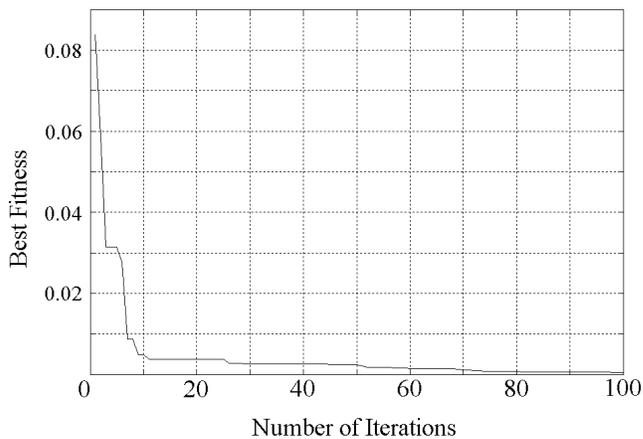

Figure 8.  Variation of best fitness with iterations for fractional order controller

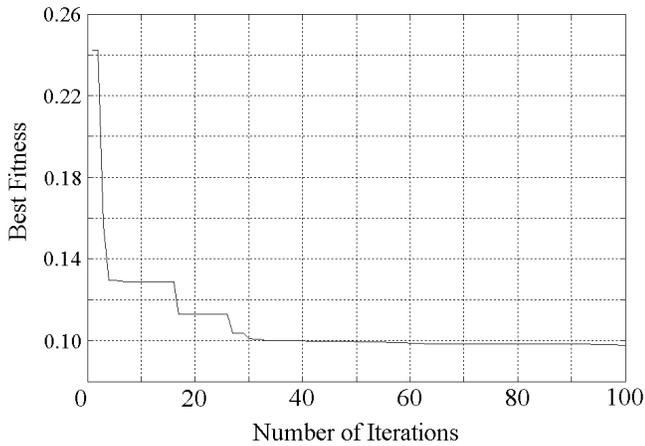

Figure 9.  Variation of best fitness with iterations for integral order controller

## V. CONCLUSIONS

It has been demonstrated that the tuning of integer order PID and fractional order $PI^\lambda D^\delta$ controllers using the proposed scheme is highly effective. The superiority of the fractional order controller is also displayed.

In the future, we plan to use fitness functions with other more stringent performance indices, and also research tuning of controllers with other stochastic optimization algorithms such as differential evolution and bacterial foraging optimization.